\documentclass[aps,superscriptaddress,preprintnumbers,showpacs,twocolumn]{revtex4}
\usepackage{amssymb}

\usepackage{indentfirst}

\usepackage{epsfig}

\usepackage{epsf}

\usepackage{graphicx}

\def\Vec#1{{\bf #1}}

\newcommand{\kp}{\mathbf{k}_\perp}

\newcommand{\p}{\perp}

\begin{document}

\newcommand*{\usm}{Center of subatomic studies and Departamento de F\'isica, Universidad T\'ecnica Federico
Santa Mar\'\i a, Casilla 110-V, Valpara\'\i so,
Chile}\affiliation{\usm}
\newcommand*{\pku}{School of Physics and MOE Key Laboratory of Heavy Ion Physics, Peking University, Beijing 100871, China}\affiliation{\pku}

\title{Double transverse spin asymmetry in the
$p^\uparrow\bar{p}^\uparrow$ Drell-Yan process from Sivers functions
}

\author{Zhun Lu}
\affiliation{\usm}
\author{Bo-Qiang Ma}\email[Corresponding author. Electronic address: ]{mabq@phy.pku.edu.cn}\affiliation{\pku}
\author{Ivan Schmidt}\email[Corresponding author. Electronic address: ]{ivan.schmidt@usm.cl}\affiliation{\usm}

\begin{abstract}

We show that the transverse double spin asymmetry (DSA) in the
Drell-Yan process contributed only from the Sivers functions can be
picked out by the weighting function
$\frac{Q_T}{M^2}(\cos(\phi-\phi_{S_1})\cos(\phi-\phi_{S_2})+3\sin(\phi-\phi_{S_1})\sin(\phi-\phi_{S_2}))$.
The asymmetry is proportional to the product of two Sivers functions
from each hadron $f_{1T}^{\perp(1)}\times f_{1T}^{\perp (1)}$. Using
two sets of Sivers functions extracted from the semi-inclusive
deeply elastic scattering data at HERMES, we estimate this asymmetry
in the $p^\uparrow\bar{p}^\uparrow$ Drell-Yan process which is
possible to be performed in HESR at GSI. The prediction of DSA in
the Drell-Yan process contributed by the function $g_{1T}(x,\Vec
k_T^2)$, which can be extracted by the weighting function
$\frac{Q_T}{M^2}(3\cos(\phi-\phi_{S_1})\cos(\phi-\phi_{S_2})+\sin(\phi-\phi_{S_1})\sin(\phi-\phi_{S_2}))$,
is also given at GSI.
\end{abstract}


\maketitle

\section{Introduction}

The Sivers effect~\cite{sivers} was proposed originally to explain
the large single spin asymmetries (SSA) observed in inclusive pion
hadro-production ($p^\uparrow  p \rightarrow \pi X$) at
FNAL~\cite{fnal90}. The effect can be quantitatively described by a
$\Vec k_T$-dependent distribution named as Sivers
function~\cite{ans95,bm} $f_{1T}^\perp(x,\Vec k_T^2)$, which is the
distribution of unpolarized partons in a transversely polarized
proton. It arises from a non-trivial correlation between the nucleon
transverse spin and the intrinsic transverse momenta in the nucleon.
Despite its (naively) $T$-odd property~\cite{collins93}, Sivers
function has been proven to be non-vanishing~\cite{bhs02} due to its
special gauge-link property~\cite{collins02,belitsky,bmp03}.

Recently the SSA measured in semi-inclusive deeply inelastic
scattering (SIDIS) processes with transversely polarized targets at
HERMES~\cite{hermespre,Airapetian:2004tw,hermes05} and
COMPASS~\cite{compass,compass06}, has been shown to be interpreted
by the Sivers effect. The asymmetry is identified by the angular
dependence $\sin(\phi-\phi_S)$, where $\phi$ and $\phi_S$ denote
respectively the azimuthal angles of the produced hadron and of the
nucleon spin polarization, with respect to the lepton scattering
plane. The coexistent Collins asymmetry~\cite{collins93}, with a
angular dependence $\sin(\phi+\phi_S)$, has also been measured in
those experiments. The data on the Sivers SSA has been utilized by
different groups~\cite{anselmino05a,anselmino05b,efr05,cegmms,vy05}
to extract the Sivers functions of the proton, especially those for
the $u$ and $d$ quarks, on the basis of the generalized
factorization~\cite{Ji:2004xq,col04}. Those sets of parametrization
of the Sivers functions are qualitatively in
agreement~\cite{anselmino05c} among themselves, and were applied to
predict the Sivers SSA in various processes in the established or
planed facilities, such as the SIDIS at JLab, and the Drell-Yan
process at COMPASS, RHIC and GSI.

In this paper, we will investigate the role of the Sivers function
on the transverse double spin asymmetry (DSA) in the Drell-Yan
process. The transverse DSA has been investigated~\cite{tdsa} for
many years, and is believed to have advantage to unravel the
transverse spin property of the nucleon~\cite{bdr}, especially the
transversity distribution $h_1(x)$~\cite{rol78}. Various azimuthal
asymmetries contributed by different $\Vec k_T$-dependent
distribution functions have been analyzed and given in
Refs.~\cite{tm94} and \cite{boer}. As shown in Ref.~\cite{boer}, the
Sivers function contributes to the DSA in the Drell-Yan process
through the product $f_{1T}^{\perp}\times f_{1T}^{\perp}$. However,
this DSA is mixed with the contribution from another $\Vec
k_T$-dependent distribution function $g_{1T}(x,\Vec k_T^2)$. We will
show that through the appropriate weighting function
$\frac{Q_T^2}{M^2}(\cos(\phi-\phi_{S_1})\cos(\phi-\phi_{S_2})
+3\sin(\phi-\phi_{S_1})\sin(\phi-\phi_{S_2}))$, the asymmetry from
the Sivers function can be isolated without mixing with the
contribution from other functions. Using two sets of
parameterizations~\cite{anselmino05b,cegmms} of the Sivers functions
we calculate the double spin asymmetry from the Sivers functions in
the $p^\uparrow \bar{p}^\uparrow$ Drell-Yan process at GSI. An
asymmetry around 1 \% is predicted. The asymmetries estimated from
these two sets of Sivers functions are quantitatively different.
Therefore measuring the DSA in the Drell-Yan process can provide new
information on Sivers functions, especially their sizes. The
transverse DSA contributed by $g_{1T}(x,\Vec k_T^2)$ through the
product $g_{1T}\times g_{1T}$ can also be picked out by another
weighting function. We estimate this asymmetry by adopting a
$g_{1T}$ coming from the combination of a Lorentz invariance
relation presented in Refs.~\cite{mulders,kot96} and the
Wandura-Wilzeck approximation~\cite{wan77}.

\section{Extracting DSA contributed by the sivers functions}

The importance of the transverse-momentum distributions of quarks
for a full understanding of the structure of hadrons has been widely
recognized in the last decade~\cite{levelt,kotzinian,mulders,bm}. A
comprehensive leading-twist tree level analysis of the (spin
dependent) Drell-Yan process in terms of $\Vec k_T$-dependent
distributions has been given in Ref.~\cite{tm94}. The role of the
$T$-odd $\Vec k_T$-dependent distributions in this process has been
presented in Ref.~\cite{boer}. In the Collins-Soper
frame~\cite{cs77} the leading order unpolarized differential
cross-section for the Drell-Yan process $h_1(P_1)+h_2 (P_2)
\rightarrow \gamma^* (q) + X\rightarrow l^+(l_1)+l^- (l_2) + X$ has
the form~\cite{boer}
\begin{eqnarray}
&&\frac{d\sigma^{(0)}(h_1h_2\rightarrow l\bar{l}X)}{d\Omega
dx_1dx_2d^2\mathbf{q}_T}= \frac{\alpha^2_{em}}{3Q^2}\sum_q
e_q^2\Bigg{\{}
A(y)\mathcal{F}[f_{1}^q f_{1}^{\bar{q}}] \nonumber\\
&& +B(y)\textmd{cos}2\phi\mathcal{F}\left [(2\hat{\mathbf{h}}\cdot
\Vec p_T\hat{\Vec h}\cdot \mathbf{k}_T -\mathbf{p}_T\cdot \Vec
k_T)\frac{h_{1}^{\perp q}
h_{1}^{\perp \bar{q}}}{M_1M_2}\right ]\Bigg{\}}\nonumber \\
 \label{cs}
\end{eqnarray}
where $q$ denotes the quark flavors, the notation
\begin{eqnarray}
\mathcal{F}[f_1f_1]&=&\int d^2\mathbf{p}_\perp
d^2\kp\delta^2(\mathbf{p}_T+\Vec k_T-\mathbf{q}_T)\nonumber\\
&&\times f_1(x_1,\Vec p_T^2)f_1(x_2,\Vec k_T^2)
\end{eqnarray}
shows the convolution of transverse momenta, $Q^2=q^2$ is the
invariance mass of the lepton pair, $q_T$ is the transverse momentum
of the lepton pair, $\hat {\mathbf{h}}=\Vec q_T/Q_T$, $\phi$ is the
angle between the hadron plane and the lepton plane, and
\begin{eqnarray}
A(y)&=&\left ( \frac{1}{2}-y+y^2 \right )=\frac{1}{4}(1+\cos^2\theta),\\
B(y)&=&y(1-y)=\frac{1}{4}\sin^2\theta,
\end{eqnarray}
in the c.m. frame of the lepton pair.

The function $h_{1}^\p$ in the second line of (\ref{cs}) is the
Boer-Mulders function~\cite{bm}, the chiral-odd partner of the
Sivers function. This function has attracted a lot of
interest~\cite{gg02,bbh03,lm04} recently because it can account for
the anomalous $\cos 2 \phi$ asymmetries~\cite{na10,conway} observed
in the unpolarized Drell-Yan process, given by the second term of
Eq.~(\ref{cs}) has shown.

The leading order differential cross-section for the double
transversely polarized Drell-Yan process is~\cite{boer}
\begin{eqnarray}
\lefteqn{ \frac{d\sigma^{(2)}(h_1^\uparrow h_2^\uparrow \rightarrow
l \bar{l} X)}
     {d\Omega dx_1 dx_2 d^2{\Vec q_T^{}}}=
\frac{\alpha^2_{em}}{3Q^2}\;\sum_{q} \;\Bigg\{ \ldots }
      \nonumber\\ &&
        + \frac{A_1(y)}{2}\;|\Vec S_{1T}^{}|\;
                         |\Vec S_{2T}^{}|\;
                \cos(2\phi-\phi_{S_1}-\phi_{S_2})\;\nonumber\\ &&
         \times  \mathcal{ F}\left[\,\Vec{\hat{h}}\!\cdot \!\Vec p_T\,
                    \,\Vec{\hat h}\!\cdot \!\Vec k_T^{}\,
                    \frac{f_{1T}^{\perp q}  f_{1T}^{\perp \bar{q}}
                         -g_{1T}^q  g_{1T}^{\bar{q}}}{M_1M_2}\right]
\nonumber\\ &&
        - \frac{A_1(y)}{2}\;|\Vec S_{1T}^{}|\;
                          |\Vec S_{2T}^{}|\;
                \cos(\phi-\phi_{S_1})\;\cos(\phi-\phi_{S_2})\;
           \nonumber\\ &&
         \times {\mathcal F}\left[\,\Vec p_T^{}\!\cdot \!
                      \Vec k_T^{}\,
                    \frac{f_{1T}^{\perp q}  f_{1T}^{\perp \bar{q}}}{M_1M_2}\right]
\nonumber\\ &&
        - \frac{A_1(y)}{2}\;|\Vec S_{1T}^{}|\;
                          |\Vec S_{2T}^{}|\;
                \sin(\phi-\phi_{S_1})\;\sin(\phi-\phi_{S_2})\;
             \nonumber\\ &&
         \times {\mathcal F}\left[\,\Vec p_T^{}\!\cdot \!
                      \Vec k_T^{}\,
                    \frac{g_{1T}^q  g_{1T}^{\bar{q}}}{M_1M_2}\right]
                     \Bigg\},\label{dsa}
\end{eqnarray}
The $\ldots$ indicates the terms which will not contribute in our
analysis below, $\phi_{S_1}$ and $\phi_{S_2}$ are the angles between
$S_{1T}$, $S_{2T}$ and the lepton plane, respectively.

As shown in (\ref{dsa}), the Sivers function can contribute to the
transverse DSA through the product $f_{1T}^{\p}\times f_{1T}^{\p}$.
However this asymmetry is mixed with the asymmetry to which it
contributes another $\Vec k_T$-dependent distribution $g_{1T}(x,\Vec
k_T^2)$. The main goal of this paper is to isolate the asymmetry
contributed by the Sivers function. The starting point is the method
introduced in Ref.~\cite{kot97}, by which one integrates the
differential cross section with a proper weighting function
$W(Q_T,\phi,\phi_{S_1},\phi_{S_2})$, as follows:
\begin{eqnarray}
 &&\langle W (Q_T, \phi,\phi_{S_1},\phi_{S_2}) \rangle
\nonumber \\
&=&  \int d\phi d \phi_{S_1} d \Vec q_T^2
        \frac{d\sigma (h_1h_2\rightarrow l \bar{l} X)}
        {d\Omega dx_1 dx_2 d^2{\Vec
        q_T^{}}} \nonumber \\
        && \times W(Q_T, \phi,\phi_{S_1},\phi_{S_2}).
\end{eqnarray}
With the above weighting procedure, one can pick up the terms in
which one is interested. Besides this, one can de-convolute the
transverse momentum integration in a model independent way.

The unpolarized angular independent cross section can be picked out
by using the weighting function $1$, from Eq.~(\ref{cs}):
\begin{eqnarray}
 &&\left ( \frac{A(y) \alpha_{em}^2} {3 Q^2} \right )^{-1}\cdot\left \langle
 1 \right \rangle_{UU} \nonumber \\
 &=&4 \pi^2\sum_q e_q^2f_1^{q}(x_1)f_1^{\bar{q}}(x_2)\label{unp}
 \end{eqnarray}

We denote $W_C=\cos(\phi-\phi_{S_1})\cos(\phi-\phi_{S_2})$ and
$W_S=\sin(\phi-\phi_{S_1})\sin(\phi-\phi_{S_2})$. Given the
weighting function $\frac{Q_T^2}{M^2}W_C$ (assuming $M_1=M_2=M$,
i.e. the colliding two hadrons are nucleons), we can obtain the
following term from (\ref{dsa}):
\begin{eqnarray}
 &&\left ( \frac{A(y) \alpha_{em}^2} {3 Q^2} \right )^{-1}\cdot\left \langle
 \frac{Q_T^2}{
 M_p^2} W_C \right \rangle_{TT} \nonumber \\
 &=&\pi^2\sum_q e_q^2 \left \{ 3\left [f_{1T}^{\p (1) q}(x_1)f_{1T}^{\p
 (1) \bar{q} }(x_2)-\right . \right . \nonumber \\
 && \left .\left .g_{1T}^{(1)q}(x_1)g_{1T}^{
 (1)\bar{q}}(x_2)\right]
 -2 f_{1T}^{\p (1) q}(x_1)f_{1T}^{\p
 (1) \bar{q}}(x_2)\right \}\nonumber\\
 &=&\pi^2\sum_a e_a^2 \left [f_{1T}^{\p (1)q}(x_1)f_{1T}^{\p
 (1) \bar{q}}(x_2)-\right . \nonumber \\
 && \left .
 3 g_{1T}^{(1) q}(x_1)g_{1T}^{
 (1) \bar{q}}(x_2)\right],\label{transint1}
\end{eqnarray}
where $f_{1T}^{\p (1)}(x)$ and $g_{1T}^{ (1)}(x)$ are the first
$\Vec k_T^2$-moments, defined as:
\begin{eqnarray}
f_{1T}^{\p (1)}(x)&=&\int d^2 \Vec k_T \frac{\Vec
k_T^2}{2M^2}f_{1T}^{\p}(x,\Vec k_T^2),\\
g_{1T}^{(1)}(x)&=&\int d^2 \Vec k_T \frac{\Vec
k_T^2}{2M^2}g_{1T}(x,\Vec k_T^2).
\end{eqnarray}
The factor $Q_T^2$ introduced in the weighting function ensures that
the transverse momentum integration in (\ref{transint1}) can be
de-convoluted (for details, refer to the Appendix). Again, applying
the weighting function $\frac{Q_T^2}{M^2}W_S$ on (\ref{dsa}), we
arrive at
\begin{eqnarray}
 &&\left ( \frac{A(y) \alpha_{em}^2} {3 Q^2} \right )^{-1}\cdot
  \left \langle \frac{Q_T^2}{
 M_p^2}W_S \right \rangle_{TT} \nonumber \\
 &=&-\pi^2\sum_q e_q^2 \left \{ 3\left [f_{1T}^{\p (1) q}(x_1)f_{1T}^{\p
 (1) \bar{q} }(x_2)-\right . \right . \nonumber \\
 && \left .\left .g_{1T}^{(1)q}(x_1)g_{1T}^{
 (1)\bar{q}}(x_2)\right]
 -2 g_{1T}^{ (1) q}(x_1)g_{1T}^{
 (1) \bar{q}}(x_2)\right \}\nonumber\\
 &=&\pi^2\sum_a e_a^2 \left [-3f_{1T}^{\p (1)q}(x_1)f_{1T}^{\p
 (1) \bar{q}}(x_2)+\right . \nonumber \\
 && \left .
  g_{1T}^{(1) q}(x_1)g_{1T}^{
 (1) \bar{q}}(x_2)\right],\label{transint2}
\end{eqnarray}

Therefore, combining (\ref{transint1}) and (\ref{transint2}), we can
extract the term contributing to the transverse DSA and coming only
from the Sivers functions:
\begin{eqnarray}
 &&\left ( \frac{A(y) \alpha_{em}^2} {3 Q^2} \right )^{-1}\cdot \left \langle \frac{Q_T^2}{
 M_p^2}\left (W_C
 +3W_S\right ) \right \rangle_{TT} \nonumber \\
 &=&-8\pi^2 \sum_q e_q^2 f_{1T}^{\p (1) q}(x_1)f_{1T}^{\p  (1) \bar{q} }(x_2),
\end{eqnarray}
with the weighting function $\frac{Q_T^2}{M^2}(W_C+3W_S)$.

By taking the ratio between (\ref{dsasivers}) and (\ref{unp}), we
define the weighted double spin asymmetry as follows
\begin{eqnarray}
A_{TT}^{f}&=&\frac{ \left \langle \frac{Q_T^2}{M^2}(W_C+3W_S) \right
\rangle_{TT}}{\langle 1 \rangle_{UU} }\nonumber \\
 &=&-\frac{ 2 \sum_q e_q^2 f_{1T}^{\p
(1)q}(x_1)f_{1T}^{\p (1)\bar{q}}(x_2)}{\sum_q e_q^2
f_1^{q}(x_1)f_1^{\bar{q}}(x_2)}. \label{dsasivers}
\end{eqnarray}

The above equation thus provides a possibility to study the Sivers
function by measuring the transverse DSA in the Drell-Yan process.

Also, from (\ref{transint1}), (\ref{transint2}) and (\ref{unp}) we
can get another type of DSA:
\begin{eqnarray}
A_{TT}^{g}&=&\frac{ \left \langle \frac{Q^2_T}{M^2}(3W_C+W_S) \right
\rangle_{TT}}{\langle 1 \rangle_{UU} }\nonumber \\
 &=&-2\frac{ \sum_q e_q^2
g_{1T}^{(1)q}(x_1)g_{1T}^{(1) \bar{q}}(x_2)}{\sum_q e_q^2
f_1^{q}(x_1)f_1^{\bar{q}}(x_2)},\label{dsag}
\end{eqnarray}
which is contributed only by $g_{1T}$.

\section{numerical results}

In this section we will give numerical results on the DSA from the
Sivers functions. We consider the transversely polarized proton
antiproton Drell-Yan process, where the valence Sivers functions are
involved, so that a larger asymmetry should be measured compared to
the $p^\uparrow p^\uparrow$ Drell-Yan process. The $p^\uparrow
\bar{p}^\uparrow$ Drell-Yan process is possible to be performed in
the planned high energy storage ring (HESR)~\cite{pax} at GSI. We
study the transverse DSA at GSI from the Sivers functions, based on
Eq.~(\ref{dsasivers}). For this end we need to know the input for
the Sivers functions. Several groups~\cite{anselmino05b,cegmms,vy05}
have parameterized the Sivers functions based on the data of SIDIS
at HERMES~\cite{Airapetian:2004tw,hermes05}, and partially based on
 COMPASS data~\cite{compass}. The kinematics in GSI can be chosen
as the c.m. energy $s=45~\textrm{GeV}^2$. For the invariance mass
square of the lepton pair we choose $Q^2=2.5~ \textrm{GeV}^2$, which
is close to the scale at HERMES. Therefore these sets of Sivers
functions extracted from the data of HERMES can be applied to
predict the asymmetries at GSI in the kinematics regime we give
above. We will adopt two sets of Sivers functions, which are the
sets in Refs.~\cite{anselmino05b} and \cite{cegmms}, respectively.
The Sivers functions in Ref.~\cite{vy05} can not be applied here
since in that paper $f_{1T}^{\p (1/2)}(x)$ is given while we use
$f_{1T}^{\p(1)}(x)$ in our calculation.

To use these Sivers functions one should notice that $T$-odd
distribution functions in the DIS and in the Drell-Yan process have
a minus sign difference~\cite{collins02}. However in the $p^\uparrow
\bar{p}^\uparrow$ Drell-Yan process two Sivers functions appear in
the product, therefore the sign difference doesn't matter here and
the functions can be used directly.

\begin{figure}

\begin{center}
\scalebox{0.85}{\includegraphics{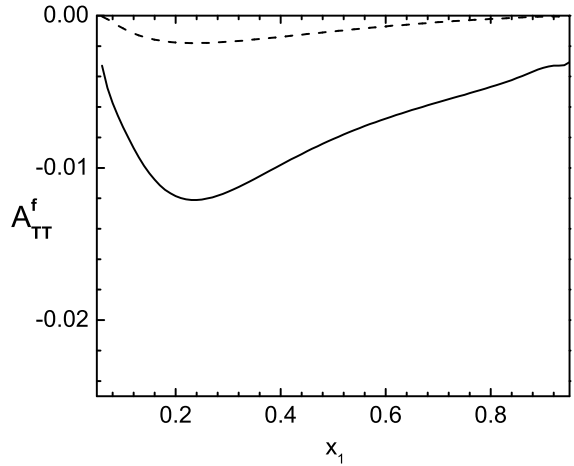}} \caption{\small The
DSA in the proton antiproton Drell-Yan process at GSI coming only
from the Sivers functions, and calculated from
Eq.~(\ref{dsasivers}). The kinematics is $s= 45~ \textrm{GeV}^2$ and
$Q^2=2.5 ~\textrm{GeV}^2$. The solid and dashed curve use the Sivers
functions in Ref.~\cite{anselmino05b} and in Ref.~\cite{cegmms},
respectively.} \label{dsaf1t}
\end{center}

\end{figure}

In Ref.~\cite{anselmino05b} the Sivers functions are parameterized
as
\begin{eqnarray}
-\frac{\Vec k_T}{M}f_{1T}^{\p,q}(x,\Vec k_T^2)=N_q (x)f_1^q(x)g(\Vec
k_T^2)h(\Vec k_T^2),
\end{eqnarray}
with
\begin{eqnarray}
N_q(x)&=&N_qx^{a_q}(1-x)^{b_q}\frac{(a_q+b_q)^{(a_q+b_q)}}{a_q^{a_q}b_q^{b_q}},\\
g(\Vec k_T^2)&=&\frac{e^{-\Vec k_T^2 /\langle k_T^2 \rangle}}{\pi
\langle k_T^2 \rangle},
\end{eqnarray}
for $q=u,d$. For the function $h(\Vec k_T^2)$ two options are
considered:
\begin{equation}
(a)~~h(\Vec
k_T^2)=\frac{2k_TM_0}{k_T^2+M_0^2},~~~(b)~~\sqrt{2e}\frac{p_T}{M^\prime}e^{-k_T^2/M^\prime}.\label{hfunction}
\end{equation}
In our calculation we will adopt option (b) in Eq.(\ref{hfunction}),
and the central values of their fit. This parametrization has taken
advantage of the more precise data~\cite{hermes05} at HERMES.

In Ref.~\cite{cegmms} the authors give the set of Sivers functions
for the $u$ and $d$ quark as,
\begin{eqnarray}
xf_{1T}^{\p(1),u}(x)=-xf_{1T}^{\p(1),d}=-0.17x^{0.66}(1-x)^5,\label{ncsiv}
\end{eqnarray}
extracted from the published HERMES data~\cite{Airapetian:2004tw},
and whose form is based on the limit of a large number of colours
$N_c$.

\begin{figure}

\begin{center}
\scalebox{0.85}{\includegraphics{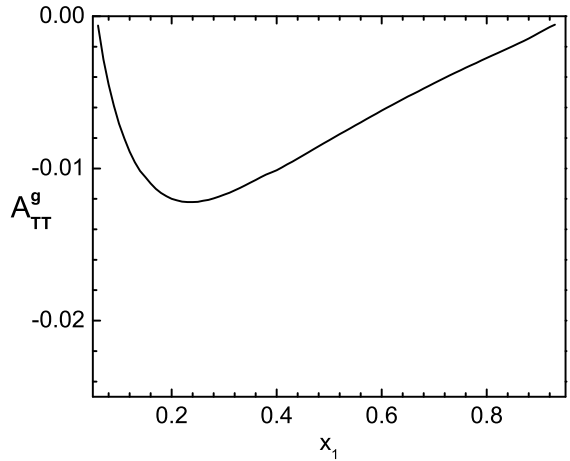}} \caption{\small The
DSA in the proton antiproton Drell-Yan process at GSI coming from
the function $g_{1T}$, defined in Eq.~(\ref{dsag}). The kinematics
is $s= 45~ \textrm{GeV}^2$ and $Q^2=2.5~ \textrm{GeV}^2$.}
\label{dsag1t}
\end{center}

\end{figure}


For the unpolarized distribution we use the MRST2001(LO set)
parametrization~\cite{mrst2001}. In Fig.~\ref{dsaf1t} we present the
DSA from Sivers functions at GSI, as a function of $x_1$. A sizable
asymmetry is predicted. The asymmetry (solid line) based on the
Sivers functions from Ref.~\cite{anselmino05b} is much larger than
the asymmetry (dashed line) based on the Sivers functions from
Ref.~\cite{cegmms}. As explained in Ref.~\cite{efremov}, taking into
account the more precise data~\cite{hermes05} of HERMES, larger
Sivers functions can be extracted compared to the parametrization in
Eq.~(\ref{ncsiv}), which will lead a larger asymmetry compared to
the dashed curve in Fig.~\ref{dsaf1t}. Thus the difference between
the asymmetries from the two sets of Sivers functions may be
reduced. 
Depending on the accuracy of the experimental measurements on the
transverse DSA at GSI, useful constraints on the Sivers functions
could be obtained, but it might be hard to distinguish between
different parameterizations without high precision measurements.

 Finally, we will predict the DSA contributed by
the function $g_{1T}(x,\Vec k_T^2)$ at GSI. This function,
describing longitudinal polarization of quarks in the transversely
polarized target, also plays role in the double polarized
(longitudinal-transverse) SIDIS process~\cite{kot96,kot06}. A
treatment on $g_{1T}(x,\Vec k_T^2)$ is the so-called Lorentz
invariance relation that connect the first $\Vec k_T^2$ moment of
$g_{1T}(x,\Vec k_T^2)$ with the twist-three distribution function
$g_2(x)$:
\begin{equation}
g_2^q(x)=\frac{d}{dx}g_{1T}^{q(1)}(x).
\end{equation}
Using the Wandzura and Wilczek approximation for $g_2^q$
\begin{equation}
g_2^q(x)\approx-g_1^a(x)+\int_x^1 d y \frac{g_1^q(x)}{y},
\end{equation}
the following relation was derived in Ref.~\cite{kot96}
\begin{equation}
g_{1T}^{(1)q}(x)\approx x\int_x^1 d y \frac{g_1^q(x)}{y}.
\end{equation}
For the polarized parton distribution we apply the GRSV2001
(standard scenario) parametrization~\cite{grsv2001}, and for the
unpolarized distribution we use GRV98 LO
parametrization~\cite{grv98}, following the choice in
Ref.~\cite{kot06}. In Fig.~\ref{dsag1t} we show the DSA contributed
by $g_{1T}(x,\Vec k_T^2)$ in the $p^\uparrow \bar{p}^\uparrow$
Drell-Yan process at GSI with $s= 45~ \textrm{GeV}^2$ and $Q^2=2.5~
\textrm{GeV}^2$. An asymmetry of $1 \%$ is predicted.

We end this section with some comment. In our calculation,
especially in the case of Siver DSA, we choose
$Q^2=2.5~\textrm{GeV}^2$. This value is consistent with the averaged
scale $\langle Q^2 \rangle$ in the HERMES experiment, from which the
Sivers functions were extracted. Therefore, the parameterizations
for Sivers functions in Refs.~\cite{anselmino05b,cegmms} can be
applied here without further assumptions. Experimental measurements
at GSI can also cover the continuous Drell-Yan masses $2 -
5~\textrm{GeV}$ which corresponds to $Q^2$ in the range $4 -
25~\textrm{GeV}^2$. To estimate the asymmetries in this region one
should use the fitted functions evolved to the relevant scale, which
is not trivial for the $\Vec k_T$-dependent
distributions~\cite{hbm02}. Therefore we assume that the ratios in
Eqs.(\ref{dsasivers}) and (\ref{dsag}) scale with $Q^2$. In this
region, The result is similar to the one which can be obtained at
the fixed value of $Q^2 = 2.5~\textrm{GeV}^2$. Also there is the
subtlety that the next to leading order correction of the hard
process could lead the substantial $K$-factor on the transversely
polarized cross-section. Since we calculate an asymmetry, which is
essentially a ratio where the $Q^2$ dependences in the numerator and
denominator tend to cancel each other, the effect of both the $Q^2$
dependence and K-factors do not introduce a strong influence on the
resulting prediction coming from Eqs.~(\ref{dsasivers}) and
(\ref{dsag}).

\section{summary}

We have performed an analysis of the transverse DSA in the Drell-Yan
process contributed by the Sivers functions through the term
$f_{1T}^{\perp}\times f_{1T}^{\perp}$. The asymmetry can be isolated
through the appropriate weighting function
$\frac{Q_T^2}{M^2}(\cos(\phi-\phi_{S_1})\cos(\phi-\phi_{S_2})
+3\sin(\phi-\phi_{S_1})\sin(\phi-\phi_{S_2}))$, without mixing with
the contribution from other distribution functions. Using two sets
of Sivers functions parameterizing the SSA data in the SIDIS
process, we calculate the double spin asymmetry in the $p^\uparrow
\bar{p}^\uparrow$ Drell-Yan process from the Sivers functions at
GSI. An asymmetry around to 1 \% is predicted. The asymmetries
estimated from these two sets of Sivers functions are quantitatively
different. Therefore measurements of the DSA in Drell-Yan process
can provide new information on the Sivers functions, especially
their sizes. The transverse DSA contributed by $g_{1T}(x,\Vec
k_T^2)$ through the product $g_{1T}\times g_{1T}$ in the Drell-Yan
process can also be picked out by a weighting function. We estimate
this asymmetry at GSI by adopting $g_{1T}$ from the combination of
the Lorentz invariance relation and the Wandura-Wilzeck
approximation. The investigation on the double transversely
polarized Drell-Yan process thus can shed light on the knowledge of
$\Vec k_T$-dependent distribution functions, including the Sivers
functions.

\begin{acknowledgments}
This work is partially supported by National Natural Science
Foundation of China (Nos.~10421503, 10575003, 10505011, 10528510),
by the Key Grant Project of Chinese Ministry of Education
(No.~305001), by the Research Fund for the Doctoral Program of
Higher Education (China), by Fondecyt (Chile) under Project
No.~3050047.
\end{acknowledgments}

\appendix*
\section{moments}
To derive (\ref{transint1}) and (\ref{transint2}) we have used the
following transverse momentum integrations:
\begin{eqnarray}
&&\int d^2 \Vec k_T d^2 \Vec p_T \delta^2(\Vec q_T-\Vec k_T - \Vec
p_T)\frac{Q^2_T}{M^2} (\Vec k_T \cdot \Vec p_T) \nonumber\\
&&\times f(x_1,\Vec
k_T^2)f(x_2,\Vec p_T^2)\nonumber\\
&=&\frac{1}{M^2}\int d^2 \Vec k_T d^2 \Vec p_T  (\Vec k_T + \Vec
p_T)^2 \Vec k_T \cdot \Vec p_T \nonumber\\
&&\times f(x_1,\Vec
k_T^2)f(x_2,\Vec p_T^2)\nonumber\\
&=&\frac{2}{M^2}\int d^2 \Vec k_T d^2 \Vec p_T  (\Vec k_T \cdot \Vec
p_T )^2 f(x_1,\Vec
k_T^2)f(x_2,\Vec p_T^2)\nonumber\\
&=&\frac{2}{M^2}\int d^2 \Vec k_T d^2 \Vec p_T \left ({\Vec k_T^1}^2
{\Vec p_T^1}^2+{\Vec k_T^2}^2 {\Vec p_T^2}^2 \right
)\nonumber\\
&&\times f(x_1,\Vec k_T^2)f(x_2,\Vec p_T^2)\nonumber\\
&=&4M^2f^{(1)}(x_1)f^{(1)}(x_2).
\end{eqnarray}
\begin{eqnarray}
 &&\int d^2 \Vec k_T d^2 \Vec p_T
\delta^2(\Vec q_T-\Vec k_T - \Vec p_T)\frac{Q^2_T}{4M^2}\hat{\Vec
h}\cdot \Vec k_T \hat{\Vec h}\cdot \Vec p_T \nonumber\\
&&\times f(x_1,\Vec
k_T^2)f(x_2,\Vec p_T^2)\nonumber\\
 &=& \frac{1}{M^2}\int d^2 \Vec k_T d^2 \Vec p_T (\Vec k_T+ \Vec p_T) \cdot \Vec k_T  (\Vec k_T+ \Vec p_T)\cdot
\Vec p_T  \nonumber\\
&&\times f(x_1,\Vec
k_T^2)f(x_2,\Vec p_T^2)\nonumber\\
&=& \frac{1}{M^2}\int d^2 \Vec k_T d^2 \Vec p_T( \Vec p^2_T \Vec
k^2_T+(\Vec k_T \cdot \Vec p_T)^2) \nonumber\\
&&\times  f(x_1,\Vec k_T^2)f(x_2,\Vec p_T^2)\nonumber\\
&=&6M^2f^{(1)}(x_1)f^{(1)}(x_2)
\end{eqnarray}
In the above integrals, the terms containing odd numbers of $\Vec
k_T^i$ or $\Vec p_T^i$ vanish after being integrated over $\Vec k_T$
or $\Vec p_T$.

\end{document}